\documentclass[conference,final,twoside]{IEEEtran}
\usepackage[T1]{fontenc}
\usepackage{microtype}
\usepackage{needspace}
\usepackage{amsmath,amssymb}
\usepackage{array,tabularx,booktabs}
\usepackage[table]{xcolor}
\usepackage{listings}
\usepackage{caption}
\usepackage[hyphens]{url}
\usepackage[hidelinks]{hyperref}
\hypersetup{pdftitle={C-rusted: The Advantages of Rust, in C, without the Disadvantages},pdfauthor={Roberto Bagnara, Abramo Bagnara}}
\newcommand{\Crusted}{\mbox{C-rusted}}
\newcommand{\Analyzer}{\emph{\Crusted{} Analyzer}}
\definecolor{cword}{HTML}{A12635}
\definecolor{ctype}{HTML}{006651}
\definecolor{cconstant}{HTML}{894600}
\definecolor{annotation}{HTML}{006296}
\definecolor{codestate}{HTML}{704067}
\definecolor{codestring}{HTML}{7C3C21}
\definecolor{codecomment}{HTML}{405C49}

\newcommand{\papercodefont}{\renewcommand{\ttdefault}{lmtt}\ttfamily}
\lstdefinestyle{paper-C}{
  language=C,
  basicstyle=\footnotesize\papercodefont,
  columns=fullflexible,keepspaces=true,showstringspaces=false,upquote=true,
  breaklines=true,breakatwhitespace=true,tabsize=2,
  aboveskip=.7\baselineskip,belowskip=.7\baselineskip,
  keywordstyle=[1]\color{cword}\bfseries,
  deletekeywords=[1]{char,double,float,int,long,short,signed,unsigned,void},
  morekeywords=[4]{_Bool,bool,char,double,float,int,long,short,signed,
    unsigned,void,size_t,ssize_t,FILE},
  keywordstyle=[4]\color{ctype},
  morekeywords=[5]{true,false,NULL,EOF,SIZE_MAX,O_RDONLY},
  keywordstyle=[5]\color{cconstant},
  morekeywords=[2]{e_checked,e_cstr,e_errno,e_f_in,e_f_out,e_in,
    e_internal_object,e_nonnull,e_nullable,e_only,e_op,e_opt,e_out,
    e_prop,e_single,e_state,e_tpar,e_track,e_unchecked,e_unsafe,e_with},
  keywordstyle=[2]\color{annotation}\bfseries,
  morekeywords=[3]{SHAR,EXCL,KEEP,INIT,VOID,LIVE,DEAD,FINI,WIPE,
    FWIPE,UNWIPED,WIPED},
  keywordstyle=[3]\color{codestate},
  stringstyle=\color{codestring},
  commentstyle=\color{codecomment}\itshape
}
\ExplSyntaxOn
\cs_new_protected:Npn \crusted_inline:n #1
  { \lstinline[basicstyle=\normalfont\ttfamily]{#1} }
\NewDocumentCommand \code { m }
  {
    \mbox
      {
        \tl_set:Nn \l_tmpa_tl {#1}
        \tl_replace_all:Nne \l_tmpa_tl { \_ } { \char_generate:nn {95}{12} }
        \tl_replace_all:Nne \l_tmpa_tl { \# } { \char_generate:nn {35}{12} }
        \exp_args:NV \crusted_inline:n \l_tmpa_tl
      }
  }
\ExplSyntaxOff
\newcommand{\readingguidecaption}{\captionsetup{width=.97\textwidth,skip=16pt}}
\begin{document}
\title{C-rusted: The Advantages of Rust, in C, \\
  without the Disadvantages}
\author{\IEEEauthorblockN{Roberto Bagnara}
\IEEEauthorblockA{BUGSENG \& University of Parma\\Parma, Italy\\
Email: \textit{name.surname}@unipr.it}
\and
\IEEEauthorblockN{Abramo Bagnara}
\IEEEauthorblockA{BUGSENG\\Parma, Italy\\
Email: \textit{name.surname}@bugseng.com}}
\maketitle
\begin{abstract}
\Crusted{} brings static verification to existing ISO~C programs.
Programmers can annotate selected types and interfaces to express who
owns a resource, when access is exclusive and when it may be shared.
The same discipline applies to C objects, operating-system resources
and resources defined by the application. Many annotations describe
properties: named facts about a value or object that a function can
require on entry, preserve or change, and guarantee on return. These
properties capture both dynamic behavior, such as ownership transfer
that depends on a call's result, and nominal types and subtypes that
distinguish values with the same C representation.

An unmodified ISO~C toolchain translates the annotated source; the
annotations used for verification add no run-time checks. The
\Analyzer{} uses abstract interpretation to check the annotations
against one another and against the C code, accounting for every
execution the C semantics and contracts permit. With all diagnostics
enabled, a completed analysis with no findings establishes that the
analyzed functions satisfy their contracts, relying only on the
contracts of the external interfaces they use and on obligations the
programmer has explicitly taken on. The resulting
guarantees exclude classes
of logic, security and run-time errors: using an expired reference,
losing the only owner of a resource, passing uninitialized bytes to an
output stream, or combining incompatible physical quantities. A team
can establish these guarantees one interface at a time while retaining
its existing C code, binary interfaces and production compiler.
\end{abstract}

\section{Stronger guarantees for existing C software}
\label{sec:cmp-motivation}

A controller manufacturer is preparing a second generation of a product.
The hardware changes, the communications protocol gains a new message,
and the software must still accept messages from installations already
in the field. The manufacturer has a C compiler for the new processor.
It also has a driver with a long operational history, a test bench,
source-level coverage measurements, a coding standard, and engineers
who know why a particular sequence of register accesses must be
preserved. Which of these assets would a change of language improve?
Which would have to be replaced, or supported a second time?

Now consider the new message handler. A truncated message fits in
the receive buffer, just as a complete one does. If the handler then
transmits the entire buffer, it can disclose whatever was left in the
unused tail. Tests that always supply a full message exercise neither
the short read nor its consequence. The manufacturer has good reasons
both to retain the C implementation and to ask for stronger evidence
about what it transmits.

\subsection{What is worth preserving}

C lets a programmer arrange data for transfer across an interface,
put storage under explicit control, and call vendor-supplied code using
the target's established binary interface. The implementation often
provides additional facilities for interrupts, memory sections, and
access to special hardware. The programmer can see when a buffer is
acquired and released. Debugging, measurement and review procedures
operate on the resulting executable, and a supplier may support that
environment for the life of the product.

The same binary interface can convey a pointer and a count while
leaving several questions to the two parties. Is the count in bytes
or elements? Who may write through the pointer? May the receiver keep
it after returning? At a device interface the difficulty can arise
even with an integer: a handle may retain its bit pattern after the
device has been closed. In each case it is the meaning of the value
at the next operation that matters. Programmers commonly communicate
this part of the design in comments or by example; every maintainer
then has to recover the convention and apply it at the next call.

ISO~C~\cite{ISO-C-2024} specifies the meaning of C constructs and the
circumstances in which a translator must issue a diagnostic. Establishing
the lifetime of a particular object at a dereference, or the extent
of initialized data at a write, requires information about that
program. A contract can supply it: the caller establishes, for
example, that a buffer is readable for a given number of elements;
the body may rely on that fact on entry. Discharging the obligation
on the development machine is especially valuable on a target where
each additional run-time check consumes time, storage or energy.

\subsection{What ownership changes}

Rust makes a particularly effective discipline part of ordinary
programming. An assignment can transfer responsibility for a resource;
a borrow grants temporary access over a lifetime that the compiler
checks. Several readers can inspect the same object, or one borrower
can have exclusive access. A caller trying to retain a reference
longer than its owner is required to change the program before it
can be accepted as safe Rust~\cite[Ch.~4]{KlabnikNK26}.

A team may introduce this discipline in new components while
continuing to maintain an existing C system. Memory-safe roadmaps
distinguish new development from replacement of deployed
components~\cite{CISA-508c}. The Linux kernel gives a concrete
example: a Rust driver can use a safe mutex abstraction around
the kernel's C mutex operations~\cite{LinuxKernel-Rust}. For a team whose
next change cuts across existing C modules, an equally practical
question is whether it can express and check the ownership convention
at a single existing C function. The original \Crusted{}
proposal~\cite{BagnaraBS23} gave it that choice, including for resources
represented by plain integers.

\subsection{From C to C-rusted}

Many \Crusted{} annotations are \emph{property based}. A property gives
a name and a set of possible values to a fact the programmer needs.
An ownership property can identify the kind of resource for which a
handle is responsible; a length property can describe initialized
storage; a property chosen by the application can give a physical
dimension to a number. A declaration can combine several such facts,
require them at entry, or relate their values on return to the
arguments and result. The C representation stays the same. For
example, two values represented by an ordinary C integer can be
distinguished by who may close the resource each one denotes.

Figure~\ref{fig:guide-properties} gives a partial guide to reading these
annotations. The \emph{\Crusted{} Technical Specification}~\cite{C-rusted-TS}
defines the language, its built-in properties, and the interpretation of
contracts. The \Analyzer{} uses the declaration at both sides of
a call: it checks that the caller supplies the required inputs
and that the implementation delivers the promised results. A
transfer that succeeds only for one return value, or a read that
initializes only part of a buffer, gives the caller a different
set of rights on each path.

A type definition, structure member or function prototype is often
the best place to record a property. One contract on a library read
can relate the returned count to initialized storage for every
client. Other facts follow from the C code and from \Crusted{} defaults.
A function whose body is available can be verified directly. For an
external routine, the maintainer chooses a contract that describes
the implementation and checks the clients against it.

Ordinary translation preserves the C representation of a handle, the
layout of a structure, and the calling convention of an annotated
function. In their ordinary compilation form, the ownership, transfer
and dimension annotations used here generate no instructions. Where
the compiler supports them, certain annotations expand to specified
native constructs, such as nullability qualifiers. Either way, the
verification step fits alongside the existing compilation route.

\subsection{Where this fits among existing checks}

Three requests commonly arise in code review. A search for likely
defects can turn up an overlooked case. An instrumented test can show
how the program behaves on a particular target and input. To establish
that a component releases a ticket at most once \emph{for every input
satisfying its contract}, the reviewer needs a proof. Expressing the
contract gives the project a precise question to put to its verifier.

Several formal methods for C address this last request. In Frama-C,
WP supports deductive verification of ACSL contracts, and Eva uses
abstract interpretation to reason about possible program
behaviors~\cite{Frama-C-WP,Frama-C-Eva}. A functional requirement
such as agreement with a mathematical algorithm may call for a
logical specification and loop invariants. \Crusted{} starts with
information a C programmer commonly knows at an interface: who owns
the object, which return value denotes success, or which bytes a
device operation fills. These facts can be expressed in the vocabulary
of the interface itself.

The choice of vocabulary has a practical consequence for teaching.
In courses in which the first author used Frama-C, students who had
already studied Hoare logic still found it difficult to turn a program
requirement into a useful deductive specification. A programmer who
can explain why a channel must be closed on an error path already
has much of what is needed to write its ownership contract. The
same project may call on a specialist to prove the correctness of
a numerical algorithm.

Checked C likewise makes gradual adoption a first-class concern:
its checked pointer types and bounds-safe interfaces are supported
by an extension to the C compiler, using dynamic checks as well as
static reasoning~\cite{ElliottRHT18}. The 3C work shows how to divide
a retrofit between a tool that infers pointer types and bounds and
a maintainer who refactors the remaining code~\cite{MachiryKMEHH22}.
\Crusted{} keeps ordinary production compilation and places resource
and value verification in the \Analyzer{}. TypeScript provides a
familiar precedent for one aspect of this design: type information can
be introduced incrementally in an established codebase~\cite{TheTypeScriptHandbook}.

For an organization using MISRA~C the two disciplines meet at the
interface. A guideline may require a coding practice across the
project; an ownership contract can identify the component responsible
for \emph{this} resource on \emph{this} return path. The declaration
then carries information that developers would otherwise have to
reconstruct during a review.
\section{C-rusted at a glance: reading a message}
\label{sec:glance}

What ought a reviewer to ask about the following piece of C? A file is
opened for reading, space is reserved for the data and a terminating zero,
and the message is passed on to another component. All the functions are
familiar. The calls even appear in a plausible order.

\begin{lstlisting}
    int fd = open(path, O_RDONLY);
    char *buf = malloc(capacity);
    ++fd;
    ssize_t n = read(fd, buf, capacity - 1U);
    buf[n] = '\0';
    process(buf);
\end{lstlisting}

Incrementing \code{fd} changes the descriptor number. The file position
is unchanged, and the new number may name another open file. Did
\code{open} and \code{malloc} succeed? For zero \code{capacity}, the
requested byte count is \code{SIZE\_MAX}. If \code{read}
returns~$-1$~\cite{POSIX-2024-read}, the next line attempts to write
the terminator one byte \emph{before} the buffer.

\begin{figure*}[!p]
\centering
\begin{minipage}[t][.83\textheight][s]{.97\textwidth}
\definecolor{guideink}{RGB}{28,57,73}
\definecolor{guidetint}{RGB}{236,242,245}
\newcommand{\guidehead}[2]{%
  {\sffamily\small\bfseries\color{annotation}#1\par}
  \vspace{4pt}{\sffamily\Large\bfseries\color{guideink}#2\par}
  \vspace{4pt}{\color{guideink}\hrule height .8pt}\vspace{5pt}}
\newcommand{\guidestep}[2]{%
  {\sffamily\large\bfseries\color{guideink}#1\quad #2\par}\vspace{5pt}}
\newcommand{\guidepanel}[1]{%
  \begingroup\setlength{\fboxsep}{8pt}%
  \noindent\colorbox{guidetint}{%
    \begin{minipage}{\dimexpr\linewidth-2\fboxsep\relax}
      \raggedright #1%
    \end{minipage}}\par\endgroup}
\newcommand{\guidecell}[3]{%
  \begin{minipage}[t]{#1}\raggedright
  {\sffamily\bfseries\color{guideink}#2\par}\smallskip
  #3\strut\end{minipage}}
\newcommand{\guidegap}{\par\vspace{9pt plus 1fill}}
\newcommand{\guidekey}[1]{\code{#1}}
\newcommand{\guidecode}{\lstinline[basicstyle=\normalsize\papercodefont]}
\normalfont\normalsize\raggedright
\guidehead{Reading C-rusted declarations \enspace/\enspace 1}
          {Properties at a C interface}
Many C-rusted annotations describe a value or resource by its
\emph{properties}. Each property has a name, a domain of possible values
and a meaning. Several properties can describe the same C value.

\guidegap
\guidestep{1}{Read a declaration}
\guidepanel{\centering
  \guidecode|void keep_message(char *e_in(own = heap) e_cstr text);|\par}
\vspace{8pt}
\guidecell{.294\linewidth}{\code{char *}}{The C parameter is a pointer
  to characters. This is how the argument is passed to the compiled function.}%
\hfill
\guidecell{.314\linewidth}{\texttt{own = heap}}{The input carries
  ownership of a heap allocation. Passing the argument transfers the
  obligation to release or pass on that allocation.}%
\hfill
\guidecell{.314\linewidth}{\code{e\_cstr}}{The input must also be a
  C string: its readable characters extend through a terminating zero.
  The caller establishes both requirements.}

\guidegap
\guidestep{2}{Find where the property must hold}
\begin{tabularx}{\linewidth}{@{}l l >{\raggedright\arraybackslash}X@{}}
\toprule
\textsf{Annotation} & \textsf{Example clause} & \textsf{How to read it}\\
\midrule
\code{e\_with} & \guidekey{own = channel} &
  A type constraint: values of the annotated handle type carry this
  ownership kind.\\[5pt]
\code{e\_in} & \guidekey{own = heap} &
  An entry requirement: the caller supplies the owner;
  the callee receives it.\\[5pt]
\code{e\_out} & \guidekey{length = 1} &
  A return guarantee: for a forward buffer, the caller may then read a
  one-element prefix. The callee establishes the promised frontier.\\
\bottomrule
\end{tabularx}\par

\guidegap
\guidestep{3}{Combine independent facts}
\guidepanel{\centering\guidecode|e_prop(fd, bool, false)|\par}
\vspace{7pt}
Here the POSIX model declares a Boolean property named \code{fd}; its
default is false. A successful acquisition establishes \code{fd} and
the corresponding ownership. The distinction shows up at the next call:
\par\vspace{7pt}
\guidecell{.475\linewidth}{Using a descriptor}{A read operation
  requires the descriptor property. The caller keeps its responsibility
  for eventually closing the descriptor.}%
\hfill
\guidecell{.475\linewidth}{Closing a descriptor}{The release operation
  also needs the ownership right. In the model used here, a successful
  close discharges the caller's obligation.}

\guidegap
\guidestep{4}{Choose what a borrower may do}
\guidecell{.475\linewidth}{\code{ref = SHAR} \enspace Shared}{Several
  readers can borrow the same resource together. The resource is protected
  against conflicting writes while a shared loan is live.}%
\hfill
\guidecell{.475\linewidth}{\code{ref = EXCL} \enspace Exclusive}{One
  borrower may modify the resource. Conflicting accesses through another
  path are excluded for the duration of the loan.}
\par\vspace{6pt}
In both cases ownership stays with its current owner. The reference kind describes
temporary access; the property \code{own} describes who must account for
the resource. Under the built-in defaults, a plain typed pointer
parameter gives the callee a one-element view: shared for a
\code{const}-qualified pointee, exclusive for a mutable one.
\end{minipage}\par
\readingguidecaption
\caption{A partial guide to the property-based annotations used in this
paper. The examples distinguish the C representation, the resource rights
attached to it, and the point at which an interface requires or establishes
a property.}
\label{fig:guide-properties}
\end{figure*}
\clearpage

Who closes the descriptor and who releases the memory?
The parameter of \code{process} is a pointer to \code{char}. Does
the callee borrow the buffer or take ownership of it? Does it expect
a terminated string? Both decisions matter at the call, yet the pointer
has the same machine representation whatever the answers.

The ordinary C compiler accepts this fragment in its enclosing function.
With the POSIX interfaces modeled, the \Analyzer{} identifies both ends
of the descriptor error. At the increment it reports an invalid operation
on the descriptor property. At the subsequent call to \code{read} it
reports that the argument no longer meets the descriptor requirement.
One finding locates the operation that lost the meaning of the integer;
the other identifies the interface that relies on it. The same run
also reports possible violations elsewhere. For example, the increment
may discard the only owner of the descriptor, \code{read} may receive a
null buffer or an oversized count, the terminator may be stored through
an invalid pointer, and the function may return still owning the
allocation.

Let us give the code an actual specification. A service receives an
already open descriptor from its caller and wants one eight-byte record.
The recipient of a complete record keeps the storage; incomplete records
are discarded. A short read is therefore a failure for this service.
A service that assembles a record across several reads needs a
different policy, as Section~\ref{sec:returns} illustrates. The caller
keeps the descriptor in either case. How can the service hand the
buffer to another component while leaving the descriptor with its caller?

\begin{lstlisting}
void keep_message(
    char *e_in(own = heap) e_cstr text);
\end{lstlisting}

Read this declaration from the caller's point of view, using
Figure~\ref{fig:guide-properties} as a key. The first property is
\code{own}; its value, \code{heap},
identifies the resource kind. On entry (\code{e\_in}) the argument
conveys ownership of a heap allocation; \code{e\_cstr} requires a
string. A pointer into a local array may designate a valid string
but confer no heap ownership. Fresh heap storage, in turn, can be
owned before it contains a string. The recipient can retain the
allocation, transfer it or release it. If it merely borrowed the
string, ownership and eventual cleanup would remain with the caller.
The header tells the reader which party assumes that obligation.

Here is the body that one might write after correcting the descriptor
arithmetic but before attending to those contracts. It is ordinary C
after preprocessing, as is its replacement below.

\begin{lstlisting}
    char *buf = malloc(9);
    ssize_t n = read(input, buf, 8);
    buf[n] = '\0';
    keep_message(buf);
    return 0;
\end{lstlisting}

Two independent failures matter here. An unsuccessful allocation can
pass a null buffer to \code{read} with a request for eight bytes;
an unsuccessful read can place the terminator at a negative subscript.
The Analyzer reports \emph{possible} violations because executions
in which both operations succeed are also admitted.

The corrected function handles both failures explicitly. Its first
annotation says that \code{input} is a descriptor. The
caller lends it to this function and retains responsibility for closing
it. Its function-level clause bounds what else the function and its
callees may write: beyond their automatic variables and what each
argument's contract permits, only the region that the modeled library
calls use for \code{errno}. A project can set such bounds once, as
defaults.

\noindent\begin{minipage}{\columnwidth}
\begin{lstlisting}
static int read_message(int e_in(fd) input)
    e_f_out(rwrite = __libc_errno) {
    char *buf = malloc(9);
    if (buf == NULL)
        return -1;
    ssize_t n = read(input, buf, 8);
    if (n != 8) {
        free(buf);
        return -1;
    }
    buf[8] = '\0';
    keep_message(buf);
    return 0;
}
\end{lstlisting}
\end{minipage}\par

Now follow the three exits. At the first, allocation failed:
there is nothing to free. At the second, the buffer exists, but the call
did not establish a complete record: the function releases its buffer
and hands nothing on. This covers both $-1$ and a successful read of
fewer than eight bytes, including zero. On the remaining path, eight
bytes have been written into an allocation of nine. Position eight
is writable, and writing a zero there establishes termination. The
recipient receives the string and the obligation to dispose of the
allocation. A subsequent call to \code{free} on this path would
dispose of storage now owned by the recipient. The \Analyzer{} validates
the corrected function, including its allocation and short-read paths.
The recipient is promised a terminated string. A protocol that
requires eight nonzero characters can specify that stronger payload
requirement at the same interface.

Notice where the work was done. The checks in the body are branches an
ordinary C implementation of this policy already needs. The extra facts
about \code{malloc} and \code{read} come from their modeled interfaces;
the extra fact about \code{keep\_message} comes from its author. All
three facts are available at verification time while the descriptor and
the pointer retain their ordinary C representations. When the
implementation of an external function is available, the same declaration also tells
the Analyzer what to check in its body; once the body meets the
contract, every verified client can rely on it.

\section{References, ownership and exceptional values}
\label{sec:references}

\subsection{Which uses of an alias are compatible?}

For C programmers a pointer parameter is a familiar example of a
reference; Rust programmers will recognize the distinction between
a shared and a mutable borrow. Consider a list node with one
dynamically allocated successor, if any:

\noindent\begin{minipage}{\linewidth}
\begin{lstlisting}
struct node {
    int value;
    struct node *e_with(own = heap) e_nullable next;
};
\end{lstlisting}
\end{minipage}\par

The link names a resource and tells us who is responsible for it.
The \code{own} property on \code{next} identifies the successor as a
heap allocation; \code{e\_nullable} permits the null link that ends a
finite list. A conventional C diagram with an arrow to the next node
looks precisely the same if that arrow is a non-owning cursor. The
difference becomes observable when a node is removed. Before replacing
an owning link, the C program must transfer its ownership elsewhere
or release the successor. Removing a cursor leaves the list's
ownership intact.

The two accessors illustrate how much an interface can already say in
C. Neither needs an annotation: the defaults make the first parameter a
shared reference and the second an exclusive one.

\begin{lstlisting}
static int node_value(const struct node *p) {
    return p->value;
}
static void node_set(struct node *p, int x) {
    p->value = x;
}
\end{lstlisting}

The first only inspects the node; the second writes its field. The
list retains ownership in both cases. In ordinary C, \code{const} prevents a
direct assignment through the first parameter; another pointer to
the same node may still permit one. A shared borrow relates the
access paths: other readers can coexist, and conflicting writes must
wait until the borrow ends. An exclusive borrow permits modification
and excludes conflicting use through another path. For two pointers
into the same array, verification takes account of which elements
each access touches and which loans are still live.

A borrow can itself be borrowed. Compare the order of accesses in
the following two bodies:

\begin{lstlisting}
static int bump(void) {
    int n = 1;
    int *parent = &n;
    int *child = parent;
    *child += 1;
    return *parent;
}
\end{lstlisting}
\begin{lstlisting}
static int late_reader(void) {
    int n = 1;
    const int *view = &n;
    n = 2;
    return *view;
}
\end{lstlisting}

In \code{bump}, the child is obtained from the parent. The parent
is used again after the last use of the child, when its loan has
ended. Both pointer variables are still in lexical scope. In
\code{late\_reader}, moving the write to \code{n} ahead of the last
read through \code{view} has the opposite effect. The \Analyzer{}
rejects the write and points back to the creation of the loan.
Code holding a shared view is entitled to read stable data through
it for the duration of that loan. Saving the value in an integer
before the write would suffice, provided that this read is the last
use of the view.

Acquisition, borrowing and consumption also explain why a pointer's
status has several components. On successful \code{malloc}, the
caller owns storage of the requested capacity; it must initialize the
bytes it intends to read. A successful \code{calloc} supplies
zero-filled storage. If a field will hold a file handle, the caller
still has to acquire the file. Moving in the opposite direction,
\code{free} consumes a heap owner. The owner must wait until the
last use of an outstanding borrow before releasing the storage.
Ownership determines who can dispose of the object; the borrow
records who still depends on it.

\subsection{One representation, two alternatives}

A null pointer and a failed acquisition of an integer file descriptor
are instances of the same idea: the interface reserves one value to
represent a separate alternative. The descriptor returned by
\code{open} uses $-1$; its \Crusted{} declaration names that alternative
with \code{e\_opt(-1)}. On success the caller receives a descriptor
and an obligation to close it. On failure it receives the sentinel
alone. The null successor of the last list node illustrates a third
use: here the distinguished value signifies the normal end of a list.

The test must discriminate the alternatives chosen by the API. The
condition \code{if (fd)} would treat the valid descriptor zero as
false and the failure value $-1$ as true. Testing
\code{fd >= 0} selects the path on which \code{open} supplies an owning
descriptor; a failed call returns $-1$.
The result from \code{read} conveys more information. Excluding $-1$
establishes a byte count. The comparison with eight in the message
example then establishes that the record is complete. Both comparisons
are ordinary C, and each supplies a different fact to the Analyzer.

Optionality can be an input contract too. The C library permits
\code{free(NULL)}. A cleanup routine accepting a nullable owner
can therefore use one release operation on both paths. In contrast,
a consumer requiring a string needs the caller to exclude the null
alternative first. The API determines which bit pattern distinguishes
the alternatives, the branches select between them, and the underlying
C representation is preserved throughout.

\section{Nominal types and their operations}
\label{sec:nominal}

A C \code{typedef} can give two distinct names to the same integer
type. A subsystem may use one for file descriptors and one for byte
counts; its C compiler still accepts one in place of the other. The
MISRA~C essential type model~\cite{MISRA-C-2025} supplies a familiar
precedent for a finer distinction: an expression answering a yes/no
question has an essentially Boolean type even when its C integer
representation supports arithmetic. \Crusted{} lets an application
designer give comparable force to its own vocabulary.

The POSIX models introduce the descriptor property with this declaration:

\begin{lstlisting}
e_prop(fd, bool, false)
\end{lstlisting}

The name of the property is \code{fd}, its domain is Boolean, and
its default is false. Successful acquisition establishes the property.
An input parameter that requires it therefore distinguishes a descriptor
from an arbitrary integer of the very same C type. Here is the modeled
declaration of \code{open}, with parameter names shortened:

\begin{lstlisting}
int e_opt(-1)
    e_out(fd  = (value >= 0) ? true : fd,
          own = (value >= 0) ? fd : own)
(open)(const char *e_cstr e_nonnull path,
       int oflag, ...) e_errno;
\end{lstlisting}

Here \code{value} denotes the returned integer. For a nonnegative
result the first equation establishes the descriptor property and
the second confers ownership of the descriptor.
The path argument has its own entry requirements: it must be a
non-null pointer to a string. The final \code{e\_errno} allows the
call's effect on \code{errno}.

Return to the first example. Testing for a nonnegative result from
\code{open} establishes both properties in its contract.
Passing the successful result to \code{read} uses its
descriptor property while the caller retains ownership. A successful
\code{close} consumes ownership in this model. By contrast, incrementing
a descriptor merely produces the next representable integer. The only
reason the test for a nonnegative return from \code{open} establishes
descriptor ownership is that \emph{this operation} confers it on that
branch. The same numerical test on an arbitrary integer cannot
establish how the process acquired it.

Identity and responsibility can diverge even when each handle has a
unique owner. POSIX \code{dup} produces a second descriptor for the
same open file description. The descriptors can share a file position,
yet each successful acquisition carries a separate close obligation.
An assignment between C integers copies the number; only an operation
such as \code{dup} acquires the second handle. Release policy likewise
belongs in the interface model. The contract used here specifies the
ownership effect of a successful close. After a close error the
correct action is platform- and API-specific: by the time a retry
is attempted, the descriptor number may already name someone else's
resource.

\subsection{Permitting the operations that make sense}

Consider two temperature readings. Taking their difference is
meaningful on either the Celsius or the Kelvin scale. Dividing one
Celsius reading by another is different: the ratio of 20 to 10
changes if the same temperatures are expressed relative to absolute
zero. A ratio of absolute temperatures is independent of the choice
of unit. Celsius readings, Kelvin readings and differences on either
scale can all use C \code{double}. A degree of temperature
\emph{difference} even has the same size on the two scales. An
application nevertheless needs to distinguish a point on a scale from
a difference between two points, then specify which operations make
sense for each.

\begin{lstlisting}
e_prop(scale, (PLAIN, CELSIUS, KELVIN, DC, DK), PLAIN)
typedef double e_with(scale = CELSIUS) celsius;
typedef double e_with(scale = KELVIN) kelvin;
typedef double e_with(scale = DC) delta_c;
typedef double e_with(scale = DK) delta_k;

e_op(sub, delta_c, celsius, celsius)
e_op(sub, delta_k, kelvin, kelvin)
e_op(div, double, kelvin, kelvin)
\end{lstlisting}

Here the property has a finite set of named values, with \code{PLAIN}
as its default. Each operator declaration specifies the operation,
result type, left operand type and right operand type, in that order.
Subtracting two Celsius readings produces a temperature difference
on that scale. Dividing two Kelvin readings produces an ordinary
\code{double}. A division of two Celsius readings has no corresponding
rule: the \Analyzer{} diagnoses an invalid property operation at the
division. The programmer specifies these distinctions at the type
declarations; the C compiler uses ordinary floating-point arithmetic.

Compatibility can also express subtyping. A non-null result can be
passed to an interface that allows null. A caller with a possibly
null value must first discriminate the alternatives before passing
it to an interface that requires a live reference. For interchangeable
functions, an implementation can accept a wider set of arguments
than the interface requires; its result must meet at least the
interface's promise. Readable and writable locations require
particular care because information crosses the boundary in both
directions. If a caller can write into a cell, every value the caller
is permitted to store must also be one its owner can accept.
These checks refine the ordinary C type-compatibility requirements
while allowing libraries to offer useful variations in their
preconditions and postconditions.

For a resource passed by reference, the interface can also specify
its state when the call begins and the state in which it must be left.
Figure~\ref{fig:guide-lifecycles} provides a reading guide for the
initializer, finalizer and erasure contracts used in the next section.

\begin{figure*}[!p]
\centering
\begin{minipage}[t][.83\textheight][s]{.97\textwidth}
\definecolor{guideink}{RGB}{28,57,73}
\definecolor{guidetint}{RGB}{236,242,245}
\newcommand{\guidehead}[2]{%
  {\sffamily\small\bfseries\color{annotation}#1\par}
  \vspace{4pt}{\sffamily\Large\bfseries\color{guideink}#2\par}
  \vspace{4pt}{\color{guideink}\hrule height .8pt}\vspace{5pt}}
\newcommand{\guidestep}[2]{%
  {\sffamily\large\bfseries\color{guideink}#1\quad #2\par}\vspace{5pt}}
\newcommand{\guidepanel}[1]{%
  \begingroup\setlength{\fboxsep}{8pt}%
  \noindent\colorbox{guidetint}{%
    \begin{minipage}{\dimexpr\linewidth-2\fboxsep\relax}
      \raggedright #1%
    \end{minipage}}\par\endgroup}
\newcommand{\guidecell}[3]{%
  \begin{minipage}[t]{#1}\raggedright
  {\sffamily\bfseries\color{guideink}#2\par}\smallskip
  #3\strut\end{minipage}}
\newcommand{\guidegap}{\par\vspace{9pt plus 1fill}}
\newcommand{\guidekey}[1]{\code{#1}}
\newcommand{\guidecode}{\lstinline[basicstyle=\normalsize\papercodefont]}
\normalfont\normalsize\raggedright
\guidehead{Reading C-rusted declarations \enspace/\enspace 2}
          {The life of a resource behind a handle}
For a pointer parameter, \code{rstate} and \code{wstate} describe the
resource it designates. One records its lifecycle; the other records
whether an overwrite of the previous contents has been established.

\guidegap
\guidestep{1}{Read the two state properties independently}
\guidepanel{\centering
  {\sffamily\bfseries Resource state}\quad
  $\underbrace{\mathtt{VOID}}_{\text{uninitialized}}
    \;\xrightarrow{\text{initialize}}\;
    \underbrace{\mathtt{LIVE}}_{\text{usable}}
    \;\xrightarrow{\text{finalize}}\;
    \underbrace{\mathtt{DEAD}}_{\text{finalized}}$\par
  \vspace{10pt}
  {\sffamily\bfseries Wipe state}\quad
  $\underbrace{\mathtt{UNWIPED}}_{\text{no overwrite guarantee}}
    \;\xrightarrow{\text{wipe}}\;
    \underbrace{\mathtt{WIPED}}_{\text{previous contents overwritten}}$\par}
\vspace{6pt}
A read requires a live resource and the appropriate access permission.
A wipe can leave a readable value in place. Wiping and then finalizing
can instead leave the resource both \code{WIPED} and \code{DEAD}.

\guidegap
\guidestep{2}{Decode an exclusive borrow}
The value of \code{excl} specifies the callee's obligation through an
exclusive reference. The entries below describe a single referent; for
a segment, an initialization guarantee covers its required readable extent.
\par\vspace{6pt}
\begin{tabularx}{\linewidth}{@{}l l >{\raggedright\arraybackslash}X@{}}
\toprule
\textsf{\texttt{excl} value} & \textsf{Entry discipline} &
  \textsf{Required effect on normal return}\\
\midrule
\guidekey{KEEP}  & Readable & No lifecycle transition required; the
  borrower may modify the value.\\[4pt]
\guidekey{INIT}  & Write first & Establish \code{rstate = LIVE} and
  \code{wstate = UNWIPED}.\\[4pt]
\guidekey{VOID}  & Write first & Leave \code{rstate = VOID}
  (uninitialized).\\[4pt]
\guidekey{FINI}  & Readable & Finalize: establish \code{rstate = DEAD}.\\[4pt]
\guidekey{WIPE}  & Write first & Overwrite the previous contents:
  establish \code{wstate = WIPED}.\\[4pt]
\guidekey{FWIPE} & Readable & Overwrite and finalize: establish
  \code{wstate = WIPED} and \code{rstate = DEAD}.\\
\bottomrule
\end{tabularx}\par
\vspace{6pt}
\emph{Write first} permits uninitialized input storage: the borrower
must write before reading through the reference. Its exit obligation
comes from the selected contract.

\guidegap
\guidestep{3}{Follow a fallible initializer}
\guidepanel{\centering
  \guidecode|int open_channel(channel_t *e_in(excl = return == 0 ? INIT : VOID) out);|\par}
\vspace{7pt}
The caller passes the address of a channel-handle slot. Both alternatives
are write-first, so that slot may be uninitialized. The function's C
return value selects the guarantee:
\par\vspace{7pt}
\guidecell{.475\linewidth}{\code{return == 0} \enspace Success: \code{INIT}}{The
  slot contains a live owning channel handle. Its owner can lend it to
  a sender and must eventually close it or transfer ownership.}%
\hfill
\guidecell{.475\linewidth}{\code{return != 0} \enspace Failure: \code{VOID}}{The
  slot is uninitialized. The caller can try acquisition again using the
  same slot; sending first would read an uninitialized handle.}
\end{minipage}\par
\readingguidecaption
\caption{A partial account of exclusive-reference contracts. The diagrams
separate lifecycle from wiping; the table gives the contract values needed
to interpret the channel and erasure examples.}
\label{fig:guide-lifecycles}
\end{figure*}
\section{Resources that change state}
\label{sec:resource-lifecycles}

A communication channel has two lifetime requirements. Sending a
message requires an open channel, and whoever acquires a channel must
eventually account for it. A send before opening violates the first
requirement. Losing an open channel on an error path violates the
second even when every send was legitimate. The C variable that holds
the handle may exist long before the channel opens and long after
it closes. The resource states and exclusive access modes in
Figure~\ref{fig:guide-lifecycles} give a precise vocabulary for these
changes.

\subsection{The caller and the implementer of an initializer}

\noindent\begin{minipage}{\columnwidth}
\begin{lstlisting}[belowskip=0pt,caption={The client closes the channel whenever initialization succeeds.},label=lst:life-channel]
typedef int e_with(own = channel) channel_t;
int open_channel(
    channel_t *e_in(excl = return == 0 ? INIT : VOID) out);
int send_message(const channel_t *c);
void close_channel(channel_t c);
\end{lstlisting}
\begin{lstlisting}
static int send_once(void) {
    channel_t c;
    if (open_channel(&c) != 0)
        return -1;
    int status = send_message(&c);
    close_channel(c);
    return status;
}
\end{lstlisting}
\end{minipage}\par

Listing~\ref{lst:life-channel} uses an interface in which the caller
provides storage and opening can fail. An integer is a plausible
representation for an operating-system handle; the annotation on the
typedef makes it an \emph{owning} handle of kind \code{channel}.
Sending requires only a shared reference to the handle. Closing takes
the handle itself, transferring responsibility to the callee. These
effects follow from the declarations and apply equally to functions
written by the library's clients.

\Needspace{4\baselineskip}
The parameter \code{out} is an exclusive reference to the caller's
storage. Its \code{excl} property selects one of two access modes
according to the result of \code{open\_channel}. On a zero result the
referent is initialized (\code{INIT}); on failure it is left
uninitialized (\code{VOID}). Both modes allow the caller to supply
uninitialized storage. Any read by the callee must be preceded by a
write establishing a valid value. Inside the annotation, \code{return}
denotes the result of the function being declared.

Follow the two exits from \code{send\_once}. At the first, acquisition
failed and the caller can return immediately. At the second,
acquisition succeeded and \code{c} owns a channel. The send borrows
it. Whether sending succeeds or fails, \code{close\_channel} takes the
handle before the caller returns the saved status. Testing the result
of acquisition identifies the live handle; the caller can leave its
local variable uninitialized until then.

A small edit exposes a common error. Return immediately on a send
failure instead of saving the status and closing the channel. On that
return \code{c} still owns a channel. The \Analyzer{} reports that the
exit would discard an active owner. Another edit produces a different
diagnostic: send a message using a local handle that has been declared
but never opened. This time the Analyzer reports the violation at the
call, where a shared borrow requires a readable handle. In the first
case the repair belongs on the exit path; in the second it belongs
before the send.

The declaration also gives the implementer a verification task.
Every successful return from \code{open\_channel} must establish
initialization. Libraries differ on the state of an output after
failure. An initializer declared with \code{VOID} on failure lets the
caller supply fresh storage; a function selecting \code{KEEP} on failure
requires readable input whenever that outcome is possible. A routine
replacing an already usable handle might legitimately provide the
latter guarantee. Choosing the mode determines both which callers can
use the function and what each can rely on afterwards.

\subsection{Finalization and guaranteed erasure}

Some interfaces finalize an object \emph{in place}; destroying a mutex
is a familiar example. The \code{FINI} mode requires an exclusive
reference to a live resource and leaves it dead. The caller may still
own the underlying storage and may construct a fresh mutex there.
Compare this with \code{close\_channel}, which takes ownership of a
handle passed by value. Both interfaces implement resource disposal;
the obligations they place on their callers are different.

Suppose a resource stores a secret. Its lifetime and the fate of the
bytes containing that secret are separate questions. \Crusted{} records
lifetime in the \code{rstate} property and records guaranteed erasure in
\code{wstate}. An erase operation can leave storage live and readable;
a finalizer can end the lifetime of a resource whose old representation
remains in storage. In the following interface the caller may read the
resulting byte, its value is zero, and the implementation must guarantee
an overwrite that survives optimization:

\noindent\begin{minipage}{\columnwidth}
\begin{lstlisting}
void wipe_byte(unsigned char
    *e_in(excl = WIPE, count = 1)
     e_out(rstate = LIVE, fill(value, 0, 1)) p);
\end{lstlisting}
\end{minipage}\par

The \code{WIPE} mode requires the overwrite. The two exit properties
say that the storage is live and that its first byte is zero.
An implementation can meet both the erasure obligation and the value
postcondition with an appropriate explicit wipe. A later ordinary
write changes the wiping state of the storage it touches. For disposal
of a resource that holds a secret, \code{FWIPE} combines guaranteed
erasure with finalization: the caller receives both guarantees for
the designated storage. The extent matters. An interface promising
to erase a key stored in an array must cover every byte of that key.

\subsection{A protocol chosen by the application}

Consider a mixer with a door and a blade. Starting the blade requires
the door to be shut; opening the door requires the blade to be stopped.
Permission to modify the mixer gives access to its controls. Whether
a particular operation is permitted also depends on earlier commands,
possibly issued by another function. The following declarations make
that history available for verification. The executable driver uses
its ordinary C calls.

\Needspace{6\baselineskip}
\begin{lstlisting}[caption={Mixer operations constrain or update the state of the door and blade.},label=lst:life-mixer]
e_state(door, (OPEN, SHUT), OPEN, join_union, order_subset)
e_state(blade, (OFF, ON), OFF, join_union, order_subset)
e_internal_object(mixer)

void shut(void) e_f_out(mixer#door = SHUT);
void start(void) e_f_in(mixer#door == SHUT)
                 e_f_out(mixer#blade = ON);
void stop(void) e_f_out(mixer#blade = OFF);
void unseal(void) e_f_in(mixer#blade == OFF)
                  e_f_out(mixer#door = OPEN);
\end{lstlisting}

Listing~\ref{lst:life-mixer} introduces two properties, each with two
named values. The third argument of \code{e\_state} is the default
at program entry. The two remaining arguments specify how to combine
information from different paths: represent alternatives by their
union and order sets by inclusion. For example, if one path leaves
the door open and another leaves it shut, both states are possible at
their join. A start is permitted only on paths on which the door
is known to be shut. The macro \code{e\_internal\_object} gives the
model a name for this mixer independently of its C representation;
\code{mixer\#door} selects its door property.

The declarations read like operating instructions. \code{shut}
establishes that the door is shut; \code{start} requires that state
and establishes that the blade is on. \code{stop} turns the blade
off in the model, thereby meeting the requirement of \code{unseal}
for opening the door. A routine beginning with the door open and the
blade off can promise to restore both entry conditions after the
following sequence:

\noindent\begin{minipage}{\columnwidth}
\begin{lstlisting}
    shut();
    start();
    stop();
    unseal();
\end{lstlisting}
\end{minipage}\par

The Analyzer uses the precondition and postcondition of each primitive
to follow this cycle. The same declarations identify two different
faults. If the door is known to be open, an immediate call to
\code{start} fails its entry condition at the call. Alternatively,
suppose a routine begins with the door shut and the blade off, and
promises to leave the blade off. Starting the blade is then legal.
Returning before stopping it breaks the routine's promise, and the
diagnostic appears at its exit. A caller could otherwise rely on that
promise to open the door.

The name \code{mixer} denotes \emph{one} machine. Giving each machine
its own internal object prevents closing one door from authorizing a start
on the other. The return convention also belongs in the protocol.
If the driver can fail to shut a door, it should establish the shut
state only on success, or after an operation that confirms closure.
The client can then use that outcome as the precondition of
\code{start}.

The same technique applies to shared storage. Name a lock, describe
how acquisition and release change its state, and declare that an
exclusive borrow of a particular buffer requires that lock to be held
exclusively. Successful acquisition then establishes the condition
for writing. Release withdraws it even if the program has retained a
pointer with exactly the same address. In a multithreaded program
the protocol also expresses who holds the permission: the right to
write belongs to the execution context that acquired the lock. The
POSIX models track the state of mutexes, read--write locks and spin
locks in the same way.
\section{What comes back from a call?}
\label{sec:returns}

Two commonplace interfaces give the return value a second job.
An operation can accept responsibility for a resource
\emph{if it succeeds}; a read can initialize part of a buffer
\emph{in an amount determined by its result}. A caller can inspect
that value to decide who now owns the resource or which bytes can be
forwarded. To justify either decision the Analyzer needs a contract
relating the result to the effect.

\subsection{Who owns a ticket after an unsuccessful offer?}

Consider a queue with a nonblocking operation for offering it a ticket.
Each ticket has an integer representation. The API designer decides
that zero means acceptance. A caller that receives any other result
can retry, return the ticket to its supplier, or consume it locally.
Only one of these parties should be responsible for a given ticket.

\begin{lstlisting}
typedef int e_with(own = ticket) ticket_t;
ticket_t obtain_ticket(void);
void consume_ticket(ticket_t value);
int offer_ticket(int e_in(own =
    return == 0 ? ticket : own) value);
void notify_acceptance(void);
\end{lstlisting}

Read the declaration of \code{offer\_ticket} as two cases. For a zero return, the
argument supplies the owning handle to a \code{ticket} resource:
the operation takes it. Otherwise the \code{own} on the right
retains the incoming ownership property: the unsuccessful caller
still owns the ticket. The parameter is a plain \code{int} because a
\code{ticket\_t} parameter would take ownership on every call. The
conditional belongs in the interface:
the result is observable by the caller even if the queue is implemented
in another translation unit.

\noindent\begin{minipage}{\columnwidth}
\begin{lstlisting}
static void forget_failed_offer(void) {
    ticket_t t = obtain_ticket();
    int status = offer_ticket(t);
    if (status != 0)
        return;
    notify_acceptance();
}
\end{lstlisting}
\end{minipage}\par

On acceptance the queue owns the ticket and the client notifies the
recipient. On failure that notification must be skipped. The early
return does skip it, but also discards the only owner of the rejected
ticket. The \Analyzer{} reports a definite loss at that return. The
repair is confined to the failure branch:

\begin{lstlisting}
static void offer_or_consume(void) {
    ticket_t t = obtain_ticket();
    int status = offer_ticket(t);
    if (status != 0) {
        consume_ticket(t);
        return;
    }
    notify_acceptance();
}
\end{lstlisting}

The consumer takes responsibility for the ticket before the client
returns. Both clients notify on success. Throughout, the ticket is
represented by the same integer; its ownership changes with the
outcome of the offer.

This convention has an inexpensive but consequential design
alternative: the queue could consume the ticket even when it cannot
enqueue it, disposing of it internally. The same C prototype and the
same return codes could serve that API. A comment which says only
that the operation can fail leaves two incompatible client cleanup
policies looking equally plausible. The conditional ownership
annotation distinguishes the two policies at every call site.

\subsection{How many bytes can the output stream read?}
\label{sec:readable}

Now consider forwarding data between two streams: the message
handler of Section~\ref{sec:cmp-motivation} in miniature. The natural unit
in the program below is one byte, so the number of complete elements
returned by \code{fread} is also the number of bytes obtained. The two
versions differ in the number of bytes they ask the output stream to
read.

\noindent\begin{minipage}{\columnwidth}
\begin{lstlisting}
static size_t forward_capacity(void)
    e_f_in(rread = *)
    e_f_out(rwrite = __libc_errno) {
    unsigned char buffer[64];
    size_t n = fread(buffer, 1,
                     sizeof buffer, stdin);
    if (n == 0)
        return 0;
    return fwrite(buffer, 1, sizeof buffer, stdout);
}
\end{lstlisting}
\end{minipage}\par

Suppose the read returns 17. Passing 64 to \code{fwrite} asks it to
take bytes 0 through 63, even though only bytes 0 through 16 came
from the stream. The check for zero excludes neither 17 nor any
other short positive read. If the read returns 64, on the other
hand, the call is legitimate. A possible diagnostic at the write
therefore describes exactly what happens: 17 and 64 give two
permitted inputs with different outcomes at the same call.

\noindent\begin{minipage}{\columnwidth}
\begin{lstlisting}
static size_t forward_initialized(void)
    e_f_in(rread = *)
    e_f_out(rwrite = __libc_errno) {
    unsigned char buffer[64];
    size_t n = fread(buffer, 1,
                     sizeof buffer, stdin);
    if (n == 0)
        return 0;
    return fwrite(buffer, 1, n, stdout);
}
\end{lstlisting}
\end{minipage}\par

The corrected count is \emph{the value returned by the operation
which produced these bytes}. Testing \code{n} and then forwarding a length obtained from
some unrelated stream would lose that connection, even if both results
lay between zero and 64. Clearing the buffer first would also
initialize its tail, but sending that tail would change the message.

The declarations supplied for the two library functions make the
argument possible. Here is the core of the model of \code{fread},
with parameter names shortened:
\begin{lstlisting}
size_t e_out(value <= nmemb)
fread(void *e_in(excl = INIT, count = size * nmemb)
          e_out(length = return * size) ptr,
      size_t size e_track(size),
      size_t nmemb e_track(nmemb), FILE *stream);
\end{lstlisting}
The result never exceeds the number of elements requested. On entry
the buffer is writable for the requested number of bytes and can be
furnished uninitialized. On return the readable prefix covers the
complete elements delivered. The \code{e\_track} annotations make the
element size and count passed at the call available to the
contract. The corresponding
input to \code{fwrite} requests initialized source bytes for the given
product.

The effect clauses on the enclosing functions bound their effects as
in Section~\ref{sec:glance}; the wildcard in \code{rread = *} also
permits reads of external regions such as the standard streams. At
the call to \code{fwrite} a separate check establishes permission to
read the payload from the buffer.

\subsection{Which end of the pointer does a bound describe?}

For a pointer \code{p} to unsigned characters there are three
different questions. How far back can this view
go? How many elements can be reached forward? Of those forward
elements, which are already fit to read? The properties
\code{before}, \code{count}, and \code{length}, respectively, answer
these questions. Forward extents start at the current pointer position.
Writing within that extent also requires a reference permitting
modification.

For example, let a six-byte array have its first four bytes initialized,
and take \code{p} to point to element 2 with a view of the whole array.
Two bytes precede \code{p}; four remain from \code{p} onward; the
first two of those four are initialized. Reading \code{p[1]} reads
element 3. Reading \code{p[2]} asks for element 4: it is within
the allocation but not initialized. Reading \code{p[4]} is a different
error: it would access beyond the array. This is a position at which
C permits the pointer to be formed and forbids its dereference.

\begin{figure}[tb]
\centering
\begin{tabular}{@{}r@{\quad}*{6}{c}@{}}
\toprule
Array index & 0 & 1 & 2 & 3 & 4 & 5\\
Index from \code{p} & $-2$ & $-1$ & 0 & 1 & 2 & 3\\
\midrule
Readable? & yes & yes & yes & yes & no & no\\
 & \multicolumn{2}{c}{\scriptsize before \code{p}} &
   \multicolumn{4}{c}{\scriptsize forward extent of \code{p}}\\
\cmidrule(r){2-3}\cmidrule(l){4-7}
\end{tabular}
\caption{Bounds measured at a pointer inside an array. Here \code{p} points
to array element 2; \code{before} is 2, \code{count} is 4, and the
exact readable prefix from \code{p} has \code{length} 2.}
\label{fig:pointer-view}
\end{figure}

Figure~\ref{fig:pointer-view} has a single transition from initialized
to uninitialized storage. A view can also contain gaps.
If elements 0, 1, 3 and 4 of a five-element array are
initialized and element 2 is not, the first two are readable and the
two beyond the gap are initialized as well. The \code{length} property
summarizes such a view with two frontiers: before the first, every element is completely
readable; from the second onward, none is completely readable; between
them, the state of the individual elements matters. In this case the
frontiers are 2 and 5. Writing the missing element can make all five
readable without writing elements 3 and 4 a second time. For an array
of structures, complete readability concerns all the members. A
program can read an initialized member even while other members still
await initialization.

\subsection{A second short read, and a short write}
\label{sec:streamloop}

A caller asks the forwarder to copy up to \code{cap} bytes. The
forwarder has 64 bytes of local storage. These are two different
quantities. A third, \code{loaded}, counts the bytes that have actually
arrived. Here is the body of \code{forward\_block}.

\noindent\begin{minipage}{\columnwidth}
\begin{lstlisting}
    unsigned char buf[64];
    if (cap > sizeof buf)
        return;

    size_t loaded = 0;
    while (loaded < cap) {
        size_t n = fread(buf + loaded, 1,
            cap - loaded, stdin);
        if (n == 0)
            break;
        loaded += n;
    }
\end{lstlisting}
\end{minipage}\par

At the head of the input loop, \code{loaded} is both the next free
index and the length of the initialized prefix. Each read asks for
at most \code{cap - loaded} bytes and writes them immediately after
that prefix. If it returns seven, the prefix grows by seven. If it
returns zero, the prefix is unchanged and the loop ends. Thus the
number of initialized bytes on exit may fall short of the request,
even when the request fits the array.

\noindent\begin{minipage}{\columnwidth}
\begin{lstlisting}
    size_t sent = 0;
    while (sent < loaded) {
        size_t n = fwrite(buf + sent, 1,
            loaded - sent, stdout);
        if (n == 0)
            break;
        sent += n;
    }
\end{lstlisting}
\end{minipage}\par

The output loop maintains a different count: \code{sent} bytes of
that prefix have already been written. The next write therefore starts
at \code{buf + sent} and may read exactly the remaining
\code{loaded - sent} bytes. A positive short write advances
\code{sent} by the number actually written, leaving the rest for the
next iteration.
The standard library contracts preserve the relation between these
counts across both loops. The \Analyzer{} validates the function
while the unused tail remains uninitialized.

Now change just the request size in the output loop:

\begin{lstlisting}
        size_t n = fwrite(buf + sent, 1,
            sizeof buf - sent, stdout);
\end{lstlisting}

This version still stays within the array and can cross the end of the
\emph{data}. Suppose \code{cap} is 64, the first
read returns 17 and the next returns zero. On the first write,
\code{sent} is zero: the correct version requests 17 bytes, the
changed version 64. The last 47 are in the allocation but were never
initialized. Conversely, when 64 bytes arrive the two expressions
coincide. The Analyzer reports a \emph{possible} borrow of
uninitialized data at \code{fwrite}.

The message reader of Section~\ref{sec:glance} rejected a short record.
Here a short block is a permissible message, and its actual length must
travel with it to the output stream.
\section{Physical quantities and tagged results}
\label{sec:quantities}

Two instruments report floating-point readings: one measures voltage,
the other resistance. Their driver can document the difference with
two C typedefs for \code{double}, but the compiler still accepts a
resistance at an interface expecting a voltage. The software can also
have the opposite problem: two names are quite properly different, yet
their values belong together in an expression. Dividing a voltage by a
resistance is just such a calculation. A property-based interface can
express the check in both directions while preserving the drivers' C
types, the arithmetic and the binary layout.

\subsection{From rejecting a sum to deriving a result}

Consider a battery monitor that computes current from those readings.
A rule requiring identical dimensions on the two sides of an addition
can catch a sum of volts and ohms. Division needs a different rule:
its operands may have different dimensions, and their quotient has a
dimension of its own. Expressing that rule once is particularly useful
in an application with a growing repertoire of units.

Use four integer properties for the powers of length, mass, time and
electric current. The suffixes \code{l}, \code{m}, \code{t} and
\code{i} identify the base dimensions. Here are the declarations and
the common C representation:

\begin{lstlisting}
e_prop(dim_l, int, 0)
e_prop(dim_m, int, 0)
e_prop(dim_t, int, 0)
e_prop(dim_i, int, 0)

typedef double quantity_t;
\end{lstlisting}

Each property has default value zero. An ampere therefore needs to
specify only that the exponent of current is one. A volt has exponent
vector $(2,1,-3,-1)$; an ohm has $(2,1,-3,-2)$. The vector belongs to
the type information used by the \Analyzer{}: the instrument still
passes a single \code{double} for each sample.

\begin{lstlisting}
e_op(add,
  quantity_t e_with(dim_l = l, dim_m = m,
                    dim_t = t, dim_i = i),
  quantity_t e_in(dim_l l, dim_m m,
                  dim_t t, dim_i i),
  quantity_t e_in(dim_l == l && dim_m == m &&
                  dim_t == t && dim_i == i))

e_op(div,
  quantity_t e_with(
    dim_l = l1 - l2, dim_m = m1 - m2,
    dim_t = t1 - t2, dim_i = i1 - i2),
  quantity_t e_in(dim_l l1, dim_m m1,
                  dim_t t1, dim_i i1),
  quantity_t e_in(dim_l l2, dim_m m2,
                  dim_t t2, dim_i i2))
\end{lstlisting}

In the addition rule the second argument of \code{e\_op} describes the
result. The following arguments describe the two operands in their C
order. The first operand binds the four exponent names; the second
must match each of them. If the match succeeds, the result has the
same dimensions. Substituting a volt and an ohm satisfies three
equalities and fails the fourth: the exponent of current differs.
The Analyzer therefore rejects that particular addition.

Division binds both exponent vectors, then subtracts the denominator's
vector from the numerator's component by component. For a voltage
divided by a resistance the calculation is
\[
   (2,1,-3,-1) - (2,1,-3,-2) = (0,0,0,1).
\]
The four exponent values identify an ampere and satisfy the return
type of the second function below. Reversing the operands would give
$(0,0,0,-1)$, the reciprocal of a current, and the return type would
reveal the mistake.

\begin{lstlisting}
static void reject_mismatched_add(
    volt_t voltage, ohm_t resistance) {
    (void)(voltage + resistance);
}

static ampere_t
current_from_voltage_and_resistance(
    volt_t voltage, ohm_t resistance) {
    return voltage / resistance;
}
\end{lstlisting}

Volts and millivolts provide an interesting design choice. Their four
exponents coincide. A project that uses volts internally can convert
a sensor's millivolt readings at the input boundary, so that all
subsequent additions use the same scale. A project that keeps both
scales can instead distinguish them by an additional property and
specify the conversion as a typed operation. A calibration interval
or a positive lower bound for the denominator can likewise be part of
the interface that acquires the readings. Each fact belongs where the
programmer can justify it.

This puts the temperature example of Section~\ref{sec:nominal} in
perspective. Celsius and Kelvin readings have the same dimension;
their relationship involves an offset. The temperature example names
points and differences and specifies the operations available to
each. The electrical example derives the type of an expression
algebraically. Both use properties to preserve the distinctions that
matter to the application.

\subsection{Which member of this result is meaningful?}
\label{sec:tagged}

Consider a C interface that returns a result record. On success it
stores a measurement in one member; on failure it stores an error code
in another. Both members have type \code{int}. For compatibility with
existing callers the layout must stay exactly as it is. The API can
nevertheless say under which condition either member may be accessed:

\begin{lstlisting}
typedef struct {
    bool is_err e_track();
    e_only(!is_err, RESULT_VALUE,
           "value belongs to a successful result") int value;
    e_only(is_err, RESULT_ERROR,
           "error belongs to a failed result") int error;
} result_t;
\end{lstlisting}

Here \code{e\_track()} makes the value of the existing Boolean field
available to annotations under its field name. The Analyzer follows its
value across C assignments and tests. The \code{e\_only} annotation on
each payload supplies an access predicate, a diagnostic name chosen
by the library author and an explanation. Even in an execution in
which both payload fields happen to contain initialized integers,
the selector determines which one the client may interpret.

\begin{lstlisting}
static int wrong_arm(const result_t *r) {
    if (r->is_err)
        return r->value;
    return 0;
}

static int select_arm(const result_t *r) {
    if (r->is_err)
        return r->error;
    return r->value;
}
\end{lstlisting}

The first function tests the selector but uses it backwards: it
reads the measurement precisely when the flag identifies an error.
The Analyzer reports the library's \code{RESULT\_VALUE} diagnostic
at that access. In the second function the two return paths read the
permitted fields, and the Analyzer validates both. Placing the
condition on the declaration extends the check to every client of the
record. Each client can establish the condition using its own
control flow.

This example uses distinct structure members, so the memory occupied
by the measurement is present even when the operation has failed.
The contract expresses entitlement to interpret that member as a
measurement. A tagged union presents a related use: its selector can
identify the active arm of overlapping storage. Similar declarations
can distinguish a calibrated reading from a raw sample. A calibration
routine establishes that property on its output, and the interface
consuming the reading requires it. The routine's implementation and
its clients can then be verified against the same definition.
\section{Who may depend on the representation?}
\label{sec:trusted-boundaries}

The program that owns an open stream decides when to close it.
The C library decides how to represent that stream: where to store
its error indicator, when to allocate a buffer, and how to release
both the buffer and the underlying file. These are two distinct
authorities. One belongs to the client of an interface; the other
belongs to its implementer. A systems language must give each of them
a way to express what the other can rely on.

\subsection{The two sides of a stream}

On success, \code{fopen} returns a pointer to a \code{FILE} object.
The client uses this pointer to read, write, query the stream or close
it. Even on a system whose header exposes all the fields of
\code{FILE}, the portable way to test the error indicator is to call
\code{ferror}. Taking a copy of the structure would duplicate its
representation while leaving the library in charge of a single open
stream. MISRA~C Rule~22.5 prohibits dereferencing a pointer to
\code{FILE}~\cite{MISRA-C-2025}.

The \Crusted{} library interface describes stream ownership independently
of the object's representation. The return type of \code{fopen}
allows a null result; a non-null result carries ownership of a resource
of kind \code{file}. The corresponding declaration of \code{fclose}
is worth reading in full:

\begin{lstlisting}
int e_opt(EOF) fclose(
    FILE *e_in(own = file, own_drop)
          e_out(rstate = DEAD)
          e_single e_nonnull stream) e_errno;
\end{lstlisting}

The caller supplies a non-null pointer to one stream and transfers its
ownership to the library. The input property \code{own\_drop} records
that the passed storage is released; the output property records the
dead resource. The return type allows the exceptional value \code{EOF}
and the final annotation permits the library to set \code{errno}.
These facts matter together on the error path. A write failure detected
while flushing the stream can make \code{fclose} return \code{EOF};
ownership has already been consumed. The caller can report the error
from the return value and the available error information. Retrying
the close would pass a dead stream to an operation that requires a
live owner.

The library model also marks the type used to represent the stream:

\begin{lstlisting}
e_unsafe(C_INTERNAL_DETAIL)
typedef struct _IO_FILE FILE;
\end{lstlisting}

Here the structure tag belongs to the implementation. The argument of
\code{e\_unsafe} names the particular issue raised by referring to an
implementation detail. A client that passes the stream pointer to
\code{ferror} uses the public interface. A client that reaches into
the object to inspect the error bit takes a dependency on the marked
representation. The \Analyzer{} can distinguish two quite different
mistakes: accessing a private field through a perfectly live stream
pointer, and calling a public operation through a pointer whose
stream has already been closed. Each demands a different repair.

\subsection{Keeping a representation private}

A channel library provides a small example on both sides of the
interface. Its maintainers store the channel state in a word of flags.
Applications need to know whether the channel is active; a diagnostic
utility also needs to record the raw word. Listing~\ref{lst:life-representation}
gives each client an interface suited to its purpose.

\begin{lstlisting}[caption={The predicate interprets a private flag; the diagnostic accessor returns the flags themselves.},label=lst:life-representation]
typedef struct {
    e_unsafe(CHANNEL_DETAIL) unsigned flags;
} channel_rep;

static int is_active(const channel_rep *c) {
    e_checked(CHANNEL_DETAIL) {
        return (c->flags & 1U) != 0;
    }
}

e_unsafe(CHANNEL_DETAIL)
static unsigned exposed_flags(const channel_rep *c) {
    e_unchecked(CHANNEL_DETAIL) {
        return c->flags;
    }
}
\end{lstlisting}

Marking the member \code{flags} makes each reference to it raise the
named issue \code{CHANNEL\_DETAIL}. In the predicate \code{is\_active},
the statement marked \code{e\_checked} takes local responsibility for
that issue. The implementation relies on bit zero being the active
indicator. Its maintainer can establish that fact from the channel's
layout and review the test along with the declaration. The caller sees
a C truth value. If a later version moves the bit, the change and its
review stay within the library.

The second client has a different requirement. It wants to interpret
or save the flags themselves. In \code{exposed\_flags}, the statement
marked \code{e\_unchecked} defers the issue raised by the field
reference. The declaration of the function uses \code{e\_unsafe} to
pass the same issue to its callers. Here is a client whose entire
interaction with the representation goes through that accessor:

\begin{lstlisting}
static unsigned application_flags(const channel_rep *c) {
    return exposed_flags(c);
}
\end{lstlisting}

The call raises \code{CHANNEL\_DETAIL} in the application. A diagnostic
utility with a versioned specification for the flags can justify that
dependency at the point where it decodes the word. Applications that
only query the channel's status can continue to use the predicate
across changes to the representation. The issue follows the private
knowledge to the client that actually uses it.

The same choice arises for library implementers. The implementation
of \code{fclose} may inspect private stream flags before releasing its
buffer. It can justify that access locally; its public contract
continues to require the transfer and consumption of ownership. A
vendor that also supplies a diagnostic extension exposing the stream's
internal state can mark the extension at its declaration. The two entry
points then make different promises about the same representation.

\subsection{Maintaining the justification}

Suppose a new public library operation tests the channel's status.
The maintainer of \code{is\_active} can call it in place of the bit test.
The reference to the
private field has gone. If the old checked statement remains, the
\Analyzer{} reports that the named issue no longer arises there.
The check works in the other direction as well: a deferred issue must
appear on the enclosing function's declaration, and a marked definition
must actually defer the issue it advertises. These checks keep the
record of the dependency in step with the implementation that needs it.

The unit of interest is a source-level reference. Taking the address of
\code{exposed\_flags} to install it in a table of diagnostic functions
already refers to the marked function. Similarly, the type or presence
of a private member can become part of a client's build through a use
in an unevaluated operand. A vendor change to the member can break that
build even if the resulting executable would never load the field.

Each justification has an identifiable scope. For the active-bit
test, the reviewer establishes the correspondence between bit zero and
the state of the channel for the layout in use. The validity of the
reference used to reach the flags remains a separate obligation checked
under the ordinary resource rules. The library maintainer can thus say
which representation the implementation supports, and a caller can
tell from the declarations which parts of that knowledge it inherits.

The same annotations apply to the diagnostics built into the
\Analyzer{}. With the name of a system diagnostic, \code{e\_checked}
records that the programmer takes responsibility for that obligation
at the statement, \code{e\_unchecked} defers it, and \code{e\_unsafe}
exposes it in the declaration. Where Rust marks a block as unsafe,
\Crusted{} also names each obligation the programmer takes on. The
named and located obligations are evidence that the project can use
in its deviation process~\cite{MISRA-Compliance-2020}.
\section{Choosing a route to stronger guarantees}
\label{sec:cmp-decision}

Consider the next release of a controller already in service. Its
maintainer wants to know whether an error during a transfer can leave
a device busy. Its project lead must decide where to spend the verification
budget, which compilers the next release will use, and who will maintain
the result. The choice of language bears on all of these questions.

Table~\ref{tab:cmp-routes} compares three routes open to this project:
continue in C with its existing assurance process, add \Crusted{}
verification to selected C interfaces, or implement a component in Rust.
The ordinary C column supplies the baseline: it identifies the assets
the project already has. The other columns show where each route adds a
guarantee and what the team must do to obtain it.

\begin{table*}[t]
\centering
\caption{Three routes for a project with an existing C product.}
\label{tab:cmp-routes}
\footnotesize
\setlength{\tabcolsep}{4pt}
\renewcommand{\arraystretch}{1.14}
\begin{tabularx}{\textwidth}{@{}>{\raggedright\arraybackslash}p{.135\textwidth}
*{3}{>{\raggedright\arraybackslash}X}@{}}
\toprule
\textbf{Decision factor} & \textbf{Ordinary C}
 & \textbf{C-rusted} & \textbf{Rust} \\
\midrule
\multicolumn{4}{@{}l}{\emph{What the project already has}}\\[2pt]
Language baseline &
Select an ISO~C edition; document implementation choices and extensions. &
Keep the production C baseline; identify the contracts and target model used for verification. &
Select a Rust edition, toolchain release and target. \\
Target and portability &
Validate the source and libraries against the production C toolchain. &
Keep the toolchain; match verification to its implementation choices. &
Check the target support tier, runtime facilities and libraries for the intended configuration. \\
Translator and assurance &
Retain applicable qualification evidence for the translator and its use. &
Retain applicable C translator evidence; assess the verification tool for its intended use. &
Select a qualified toolchain where required; check its scope for the target and use. \\
Development tools &
Retain build, debugging, test and coverage workflows. &
Retain the C workflow and add static checking of the annotated code. &
Provide a Rust workflow, including debugging and coverage across any language boundary. \\
\midrule
\multicolumn{4}{@{}l}{\emph{What the project can gain, and what adoption changes}}\\[2pt]
Memory and resources &
Use design rules and verification tools to establish ownership, bounds and resource protocols. &
Check selected functions against ownership, bounds, initialization and resource contracts. &
Safe Rust enforces memory safety; types, ownership and destructors can encode resource protocols. \\
Existing code and interfaces &
Keep C code and C interfaces. &
Keep C code; describe externally supplied functions using contracts. &
Keep C components through a foreign-function interface; establish the safety of the Rust wrapper. \\
Unit of adoption &
Add tests, analyses or coding rules within existing modules. &
Start with a function or interface; extend verification to more callers and definitions. &
Introduce a Rust component alongside C, or choose Rust for a new subsystem. \\
Coding guidelines &
Apply the project's C guidelines and compliance process. &
Keep that process; supply evidence for selected obligations and name those the programmer takes on. &
Apply Rust guidance; review unsafe code and interfaces between the languages. \\
People and training &
Draw on the team's existing C and target experience. &
Train contract authors and reviewers; maintain the remaining modules as ordinary C. &
Develop Rust design and tooling skills, and mixed-language review skills where C remains. \\
\bottomrule
\end{tabularx}
\end{table*}

\subsection{Which guarantee does the interface express?}

A Rust library can expose a safe borrowing interface whose lifetimes
the compiler checks at each safe call site. If its implementation uses
unsafe operations, the library author must establish that calls admitted
by the public interface meet their safety requirements. Memory safety
permits a process to leak an allocation~\cite{RustReference-Leaks}. The
requirement to roll back a transaction, return a descriptor to its pool,
or leave a device idle comes from the application. A Rust type can
encode that policy: a guard can own the transaction, for example, and
its destructor can roll it back when the guard goes out of scope.

In C, a library can pass a descriptor as an integer and document the
protocol for that integer. A reviewer or a verification tool can then
establish whether a particular caller follows the protocol. In
\Crusted{}, the declaration itself distinguishes a callee that borrows
the descriptor from one that takes ownership. Suppose a failed call
returns ownership to the caller and the caller immediately returns an
error. Every integer operation on that path is valid C. The \Analyzer{}
reports the lost resource at the return where the last owner goes out
of scope. An evaluator can read the contract to identify the policy and
inspect the verified calls and definitions to see where it has been
established.

\subsection{Where can a project start?}

A Rust library can supply a service behind a C ABI; a Rust component can
also call an existing C library. The foreign-function interface has to
agree on representations, calling conventions, ownership and error
handling~\cite{RustNomicon-FFI}. Consider a C driver that keeps a pointer to a
callback context until the callback is unregistered. A wrapper that
allows the Rust caller to drop the context as soon as registration
returns leaves the driver with a dangling pointer. A wrapper that ties
the context's lifetime to the registration can establish that invariant
in one place for every safe client of the driver.

With \Crusted{}, a maintainer can begin by specifying the ownership
effect of a single C function. Its symbol, calling convention and
existing callers remain in place. For each caller brought under
verification, the \Analyzer{} checks the obligations imposed by the
contract. Checking the function body establishes what the function
promises in return. For a separately supplied binary, the maintainer
records which version satisfies the contract and on what evidence.
When an interface changes, rechecking its callers identifies the
sites that relied on the old promise.

Automated translation offers a third starting point for a Rust component.
Emre et al.\ distinguish translating a C pointer into a raw Rust pointer
from recovering a Rust reference whose uses the compiler can
check~\cite{EmreSDH21}. For the driver, that distinction determines
who must establish how long the callback context stays alive: each user
of a raw interface, or the author of a checked abstraction that clients
can reuse. A tool proposing a \Crusted{} contract would face the same
question about the external driver. In both cases, the driver
specification or its implementation supplies the evidence for the
lifetime asserted at its boundary.

\subsection{Which platform and assurance evidence transfer?}

For an embedded product, portability raises three separate questions.
Which language edition describes the source? Which compiler, runtime
facilities and libraries are available on the target? What evidence
does the project hold for their use in this device? A C program may
depend on an integer width, structure alignment or the signedness of
plain \code{char}. \Crusted{} retains the production C configuration;
its verification model must account for those choices. For Rust, the
published target tiers specify which combinations are built, tested
and supported~\cite{Rustc-Platform-Support}. A comparison between two candidate
configurations can therefore start with a precise target and the
facilities the application actually uses.

The draft \Crusted{} specification~\cite{C-rusted-TS} explores a candidate
notation that uses namespaced C23 attributes to express the same
contracts. Their position can identify a particular type within a
declarator. The macro notation used in this paper works with C99
and later editions.

Language specification and tool qualification address different parts
of the decision. A C project can name the ISO edition it uses and the
implementation choices the supplier documents. Qualification evidence
for its existing translator remains applicable under its stated
conditions of use. Introducing the \Analyzer{} calls for an assessment
of the reliance the project will place on its results. For Rust, the
FLS, originally the Ferrocene Language Specification, describes the
language, and the edition mechanism lets crates
using different editions interoperate~\cite{Rust-FLS,Rust-Edition-Guide}.
Ferrocene provides an example of a Rust toolchain qualified for
specified safety-related uses~\cite{Ferrocene23}. In each case the
evidence belongs to a particular configuration and intended use.

An organization applying MISRA~C:2025 gains from the relationship in
both directions. MISRA~C constrains C constructs beyond the properties
expressed in \Crusted{} contracts. In return, the \Analyzer{} can
provide evidence about the lifetime of a referenced object, the extent
available for a pointer access, the initialization of a value or the
release of a resource. The project can use those results in its wider
compliance process~\cite{MISRA-Compliance-2020}. Teams evaluating Rust can
consult MISRA's analysis of the applicability of its C guidelines to
Rust and the coding guidance of the Safety-Critical Rust
Consortium~\cite{MISRA-C-2025-Addendum-6,Rust-Coding-Guidelines}. A comparison focused
on the needs of embedded systems is given in~\cite{BagnaraVCBS25}.

\subsection{Who must learn what?}

The relevant people for a controller already in service are the people
who know its hardware, reproduce its builds, review changes and will
inherit its maintenance. A maintainer of an unchanged C module can carry
on working with the familiar tools. The person maintaining an annotated
interface learns to distinguish what a checked definition establishes
from what a contract assumes about a binary, to interpret a possible
diagnostic, and to trace the effects of a contract change through its
callers. A team adopting Rust needs engineers who can design Rust
abstractions and review their implementations. Where the product retains
C components, the team also needs expertise on both sides of the ABI.

Return to the controller's new message handler. When the handler can be
isolated behind a stable interface, a supported Rust configuration and
a single well-reviewed wrapper can serve many clients. When the change
runs across C modules tied to a vendor compiler and the immediate issue
is how many bytes a library call initialized, a \Crusted{} contract can
state that fact at the existing interface and give every verified caller
the benefit of it. A team whose existing C assurance process already
establishes the required property can retain that evidence. For this
decision, the useful boundary is the one at which the team can express
the guarantee, establish it, and keep it true across subsequent releases.
\section{The C-rusted Analyzer}
\label{sec:verification}

For the copying loops of Section~\ref{sec:streamloop}, the \Analyzer{}
establishes that every byte offered to the output stream was
initialized by an earlier read, for every permitted sequence of short
reads and writes. Connecting facts across C expressions, control flow
and library interfaces in this way is the purpose of the Analyzer.

\subsection{From a declaration to a proof about C code}

The \Analyzer{} is a Clang-based implementation of the language. It
uses abstract interpretation~\cite{CousotC77} to reason about sets
of states that can reach a program point. In the first stream example,
one such set contains all the positive byte counts up to the capacity
of the buffer. The result of the read denotes both a C integer and the
extent of the newly initialized data. Retaining that relation makes it
possible to justify the subsequent write for every count in the set.

The unit of analysis is a function body. At its entry, the declared
requirements describe the inputs. Each operation must be legitimate in
the states that reach it, and each return must establish the promised
effects. At a call site, the caller supplies the required arguments
and may use the callee's guarantees. The \Crusted{} library models
supply such contracts for the C standard library, from C99 to C23, and
for a broad subset of POSIX. For an external library, firmware routine
or device, the maintainer supplies the corresponding evidence as part
of the interface specification. Function-scoped analysis also allows
independently developed modules to be verified separately.

The analysis is conservative: it accounts for every execution that the
C semantics and the contracts permit, and classifies each violation
such an execution can exhibit as definite or possible. With all
diagnostics enabled, a completed analysis with no findings establishes
that the analyzed functions meet their contracts. The only further
premises are obligations the programmer has explicitly taken on, such
as a statement marked \code{e\_checked} (Section~\ref{sec:trusted-boundaries}).

These arguments apply to the C program as translated for its target.
For example, if a parser loads a byte into a plain \code{char}, the
target's choice of signedness may determine which branch handles a
high-bit byte. When that branch transfers ownership, signedness
matters to the proof of a resource lifetime. BUGSENG's ECLAIR Software
Verification Platform determines such
implementation-defined choices from the production compiler, target and
invocation. In its integration with the \Analyzer{}, that configuration
provides the C semantics for verification. The project's own C
toolchain continues to produce the executable.

\subsection{What a diagnostic tells the programmer}

A violation is \emph{definite} when every execution that the analysis
represents at the operation exhibits it, and \emph{possible} when the
analysis represents both violating and nonviolating executions there. On the
failure return of the faulty queue client, every state still carries
the obligation for the ticket, so the violation is definite and the
repair belongs on that path. At a write of the whole buffer after a
positive read, a full read is legitimate and a short one is not: the
violation is possible, and a read of 17 bytes shows that it can
really occur. The repair passes the count returned by the read to the
write.

More generally, a possible diagnostic identifies a proof obligation
that remains open, whether the violating case is real or arises from
the approximation. The programmer can inspect the operation and its
inputs, refine an imprecise contract, or take responsibility for the
case by name, as in Section~\ref{sec:trusted-boundaries}. Once all obligations are discharged, the contracts state
exactly what a client of that verified code can rely on.

\subsection{Related work}

Each ingredient of this design has been explored elsewhere. Microsoft's
SAL relates a buffer to the count that a call fills~\cite{HackettDWY06};
the design of Clang's \texttt{-fbounds-safety}~\cite{Clang-BoundsSafety}
and Deputy~\cite{ConditHAGN07} attach bounds to pointers. Cyclone brought
regions and tagged unions to a safe dialect of C~\cite{JimMGHCW02};
RefinedC~\cite{SammlerLKMDG21} and Cake~\cite{Cake} check ownership in
C programs. Vault enforced protocols with typestate~\cite{DeLineF01},
Clang's thread safety analysis checks lock
capabilities~\cite{HutchinsBS14}, and Osprey checked dimensional units
in C~\cite{JiangS06}. Astr\'ee showed that sound abstract
interpretation scales to industrial C code~\cite{CousotCFMMMR07}.
\Crusted{} expresses bounds, ownership, protocols and units in a single
property vocabulary over unmodified C, and checks them all with one
analyzer.

\section{C-rusted Explorer}
\label{sec:explorer}

The \Crusted{} Explorer brings the \Analyzer{} into the browser; the
\href{https://bugseng.com/c-rusted/}{\Crusted{} project page} gives
access to it. The Explorer provides a runnable version of every C
example in this paper, including the variants that illustrate a
diagnostic. Search the catalog for \emph{paper} to find them; the
\emph{Stream I/O} group, for example, reproduces the readable-prefix
argument of Section~\ref{sec:streamloop}.
A reader can open a working example, change the C code or its
annotations, run an analysis and go from a finding straight to the
source expression that produced it. Related notes accompany the
finding; a successful run explicitly reports that there are no
findings.

The \emph{Resource protocols} group puts the lock technique of
Section~\ref{sec:resource-lifecycles} to work. In \emph{Writing after
unlock}, a routine stores a sample in storage guarded by a lock, but
only after releasing the lock. Predict where the diagnostic will
appear. Then move the store between the calls that acquire and release
the lock, and compare the result with \emph{Writing with the lock
held}. The address of the storage is the same in both versions; only
the protocol state at the store has changed.

For a different experiment, choose \emph{Incompatible physical
quantities}. Change an operand so that both addends have the same
dimensions, then compare the result with \emph{Adding equal
quantities}. In \emph{Deriving current}, division computes a new
dimension from those of its operands. Such experiments let an API
designer try out an interface while its representation and production
compiler remain fixed.

\section{Conclusion}

\Crusted{} gives the maintainer of an existing C program a way to
specify distinctions on which the program already relies. An integer
can carry the obligation to close a descriptor. A count can say
how much of a buffer is ready to transmit. A state transition can
make a mixer operation available to its next caller. Each of these
facts can be expressed in a C declaration and checked against the
implementation by the \Analyzer{}.

The useful first step may be one function: choose an interface at
which a mistake would matter, state what its parties owe one another,
and verify the code on each side. The \Crusted{} Explorer, available
from \url{https://bugseng.com/c-rusted/}, is a convenient place to try
that step. The resulting contract can then serve the next caller and
the next implementation. That is how the guarantees can grow across a
codebase while the team continues to ship the C it maintains.

\section*{Acknowledgment}
We thank Federico Serafini for his contributions
to \Crusted{}. Serafini coauthored the original arXiv paper and gave an
early account of the language in his master's thesis~\cite{Serafini23}.
Simone Ballarin and Patricia M.~Hill worked with us on the first draft of
a paper on \Crusted{}. Nicola Vetrini helped write the early Technical
Specification and mapped its provisions to MISRA~C. Filippo Martelli
contributed to the specification and compared borrowing and lifetimes
through paired C and Rust examples in his master's thesis~\cite{Martelli26}.

We thank Luca Ciucci for his work with us on the comparison of C, Rust,
\Crusted{} and MISRA~C for embedded world 2025~\cite{BagnaraVCBS25},
and Lavinia Battaglia and Patricia M.~Hill for their help in revising
that manuscript.

We are grateful to Thiago Adams, Jens Gustedt, Marcus Johnson, David
Keaton, Miguel Ojeda and Martin Uecker for their comments, suggestions
and encouragement during the development of the specification. We also
thank Giuseppe Iacono, Felice Francesco Lorenzano, Francesco Lucchesi,
Gael Mouraret and Federico Tamponi for their contributions as BUGSENG
trainees.

\IEEEtriggeratref{24}

\end{document}